\documentclass{aastex631}
\usepackage{epsfig}
\usepackage{rotating}
\usepackage{natbib}

\newcommand{\be}{\begin{equation}}
\newcommand{\ee}{\end{equation}}
\newcommand{\bary}{\begin{eqnarray}}
\newcommand{\eary}{\end{eqnarray}}

\graphicspath{{./}{figures/}}

\begin{document}

\title{\Large Constraints on the very high energy gamma-ray emission from short GRBs with HAWC}

\author{A.~Albert}
\affiliation{Physics Division, Los Alamos National Laboratory, Los Alamos, NM, USA}
\author[0000-0001-8749-1647]{R.~Alfaro}
\affiliation{Instituto de Física, Universidad Nacional Aut\'onoma de México, Ciudad de Mexico, Mexico }
\author{C.~Alvarez}
\affiliation{Universidad Autónoma de Chiapas, Tuxtla Gutiérrez, Chiapas, México}
\author{J.C.~Arteaga-Velázquez}
\affiliation{Universidad Michoacana de San Nicolás de Hidalgo, Morelia, Mexico }
\author{K.P.~Arunbabu}
\affiliation{Department of Physics, St. Albert’s College (Autonomous), Cochin, 682018 Kerala, India.}
\author[0000-0002-4020-4142]{D.~Avila Rojas}
\affiliation{Instituto de Física, Universidad Nacional Autónoma de México, Ciudad de Mexico, Mexico }
\author[0000-0002-2084-5049]{H.A.~Ayala Solares}
\affiliation{Department of Physics, Pennsylvania State University, University Park, PA, USA }
\author{R.~Babu}
\affiliation{Department of Physics, Michigan Technological University, Houghton, MI, USA }
\author[0000-0003-3207-105X]{E.~Belmont-Moreno}
\affiliation{Instituto de Física, Universidad Nacional Autónoma de México, Ciudad de Mexico, Mexico}
\author{C.~Brisbois}
\affiliation{Department of Physics, Michigan Technological University, Houghton, MI, USA }
\author[0000-0002-4042-3855]{K.S.~Caballero-Mora}
\affiliation{Universidad Autónoma de Chiapas, Tuxtla Gutiérrez, Chiapas, México}
\author[0000-0003-2158-2292]{T.~Capistrán}
\affiliation{Instituto de Astronomía, Universidad Nacional Autónoma de México, Ciudad de Mexico, Mexico }
\author[0000-0002-8553-3302]{A.~Carramiñana}
\affiliation{Instituto Nacional de Astrofísica, Óptica y Electrónica, Puebla, Mexico }
\author[0000-0002-6144-9122]{S.~Casanova}
\affiliation{Institute of Nuclear Physics Polish Academy of Sciences, PL-31342 IFJ-PAN, Krakow, Poland }
\author{O.~Chaparro-Amaro}
\affiliation{Centro de Investigación en Computación, Instituto Politécnico Nacional, México City, México.}
\author{U.~Cotti}
\affiliation{Universidad Michoacana de San Nicolás de Hidalgo, Morelia, Mexico }
\author{J.~Cotzomi}
\affiliation{Facultad de Ciencias Físico Matemáticas, Benemérita Universidad Autónoma de Puebla, Puebla, Mexico}
\author[0000-0002-7747-754X]{S.~Coutiño de León}
\affiliation{Department of Physics, University of Wisconsin-Madison, Madison, WI, USA }
\author[0000-0002-8528-9573]{C.~de León}
\affiliation{Universidad Michoacana de San Nicolás de Hidalgo, Morelia, Mexico }
\author[0000-0001-9643-4134]{E.~De la Fuente}
\affiliation{Departamento de Física, Centro Universitario de Ciencias Exactase Ingenierias, Universidad de Guadalajara, Guadalajara, Mexico }
\author{R.~Diaz Hernandez}
\affiliation{Instituto Nacional de Astrofísica, Óptica y Electrónica, Puebla, Mexico }
\author{S.~Dichiara}
\affiliation{Department of Astronomy and Astrophysics, The Pennsylvania State University, 525 Davey Lab, University Park, PA 16802, USA}
\author{B.L.~Dingus}
\affiliation{Physics Division, Los Alamos National Laboratory, Los Alamos, NM, USA}
\author[0000-0002-2987-9691]{M.A.~DuVernois}
\affiliation{Department of Physics, University of Wisconsin-Madison, Madison, WI, USA }
\author[0000-0003-2169-0306]{M.~Durocher}
\affiliation{Physics Division, Los Alamos National Laboratory, Los Alamos, NM, USA }
\author[0000-0002-0087-0693]{J.C.~Díaz-Vélez}
\affiliation{Departamento de Física, Centro Universitario de Ciencias Exactase Ingenierias, Universidad de Guadalajara, Guadalajara, Mexico }
\author[0000-0001-5737-1820]{K.~Engel}
\affiliation{Department of Physics, University of Maryland, College Park, MD, USA }
\author[0000-0001-7074-1726]{C.~Espinoza}
\affiliation{Instituto de Física, Universidad Nacional Autónoma de México, Ciudad de Mexico, Mexico}
\author{K.L.~Fan}
\affiliation{Department of Physics, University of Maryland, College Park, MD, USA }
\author[0000-0002-0173-6453]{N.~Fraija $^{a}$}
\affiliation{Instituto de Astronom\'ia, Universidad Nacional Autónoma de México, Ciudad de Mexico, Mexico, \href{mailto:nifraija@astro.unam.mx}{$^{a}$nifraija@astro.unam.mx}, \href{mailto:nimagda@astro.unam.mx}{$^{b}$magda@astro.unam.mx}, \href{mailto:yfperez@astro.unam.mx}{$^{c}$yfperez@astro.unam.mx}}
\author{A.~Galván-Gámez}
\affiliation{Instituto de Astronom\'ia, Universidad Nacional Autónoma de México, Ciudad de Mexico, Mexico}
\author[0000-0002-4188-5584]{J.A.~García-González}
\affiliation{Tecnologico de Monterrey, Escuela de Ingeniería y Ciencias, Ave. Eugenio Garza Sada 2501, Monterrey, N.L., Mexico, 64849}
\author[0000-0003-1122-4168]{F.~Garfias}
\affiliation{Instituto de Astronomía, Universidad Nacional Autónoma de México, Ciudad de Mexico, Mexico }
\author[0000-0002-5209-5641]{M.M.~González $^{b}$}
\affiliation{Instituto de Astronomía, Universidad Nacional Autónoma de México, Ciudad de Mexico, Mexico, \href{mailto:nifraija@astro.unam.mx}{$^{a}$nifraija@astro.unam.mx}, \href{mailto:nimagda@astro.unam.mx}{$^{b}$magda@astro.unam.mx}, \href{mailto:yfperez@astro.unam.mx}{$^{c}$yfperez@astro.unam.mx}}
\author[0000-0002-9790-1299]{J.A.~Goodman}
\affiliation{Department of Physics, University of Maryland, College Park, MD, USA }
\author[0000-0001-9844-2648]{J.P.~Harding}
\affiliation{Physics Division, Los Alamos National Laboratory, Los Alamos, NM, USA }
\author[0000-0002-2565-8365]{S.~Hernandez}
\affiliation{Instituto de Física, Universidad Nacional Autónoma de México, Ciudad de Mexico, Mexico}
\author[0000-0002-7609-343X]{B.~Hona}
\affiliation{Department of Physics, Michigan Technological University, Houghton, MI, USA }
\affiliation{Department of Physics and Astronomy, University of Utah, Salt Lake City, UT, USA }
\author[0000-0002-3808-4639]{D.~Huang}
\affiliation{Department of Physics, Michigan Technological University, Houghton, MI, USA }
\author[0000-0002-5527-7141]{F.~Hueyotl-Zahuantitla}
\affiliation{Universidad Autónoma de Chiapas, Tuxtla Gutiérrez, Chiapas, México}
\author{T.B.~Humensky}
\affiliation{Department of Physics, University of Maryland, College Park, MD, USA }
\author{P.~Hüntemeyer}
\affiliation{Department of Physics, Michigan Technological University, Houghton, MI, USA }
\author[0000-0001-5811-5167]{A.~Iriarte}
\affiliation{Instituto de Astronomía, Universidad Nacional Autónoma de México, Ciudad de Mexico, Mexico }
\author[0000-0003-4467-3621]{V.~Joshi}
\affiliation{Max-Planck Institute for Nuclear Physics, 69117 Heidelberg, Germany}
\author{S.~Kaufmann}
\affiliation{Universidad Politecnica de Pachuca, Pachuca, Hgo, Mexico }
\author[0000-0001-6336-5291]{A.~Lara}
\affiliation{Instituto de Geofísica, Universidad Nacional Autónoma de México, Ciudad de Mexico, Mexico }
\author[0000-0002-2467-5673]{W.H.~Lee}
\affiliation{Instituto de Astronomía, Universidad Nacional Autónoma de México, Ciudad de Mexico, Mexico }
\author[0000-0001-5516-4975]{H.~León Vargas}
\affiliation{Instituto de Física, Universidad Nacional Autónoma de México, Ciudad de Mexico, Mexico }
\author{J.T.~Linnemann}
\affiliation{Department of Physics and Astronomy, Michigan State University, East Lansing, MI, USA }
\author[0000-0001-8825-3624]{A.L.~Longinotti}
\affiliation{Instituto de Astronomía, Universidad Nacional Autónoma de México, Ciudad de Mexico, Mexico }
\author[0000-0003-2810-4867]{G.~Luis-Raya}
\affiliation{Universidad Politecnica de Pachuca, Pachuca, Hgo, Mexico }
\author[0000-0001-8088-400X]{K.~Malone}
\affiliation{Space Science and applications, Los Alamos National Laboratory, Los Alamos, NM, USA }
\author{S.S.~Marinelli}
\affiliation{Department of Physics and Astronomy, Michigan State University, East Lansing, MI, USA }
\author[0000-0001-9052-856X]{O.~Martinez}
\affiliation{Facultad de Ciencias Físico Matemáticas, Benemérita Universidad Autónoma de Puebla, Puebla, Mexico }
\author[0000-0002-2824-3544]{J.~Martínez-Castro}
\affiliation{Centro de Investigación en Computación, Instituto Politécnico Nacional, México City, México.}
\author[0000-0002-2610-863X]{J.A.~Matthews}
\affiliation{Dept of Physics and Astronomy, University of New Mexico, Albuquerque, NM, USA}
\author[0000-0002-8390-9011]{P.~Miranda-Romagnoli}
\affiliation{Universidad Autónoma del Estado de Hidalgo, Pachuca, Mexico }
\author{J.A.~Morales-Soto}
\affiliation{Universidad Michoacana de San Nicolás de Hidalgo, Morelia, Mexico }
\author[0000-0002-1114-2640]{E.~Moreno}
\affiliation{Facultad de Ciencias Físico Matemáticas, Benemérita Universidad Autónoma de Puebla, Puebla, Mexico }
\author[0000-0002-7675-4656]{M.~Mostafá}
\affiliation{Department of Physics, Pennsylvania State University, University Park, PA, USA }
\author[0000-0003-0587-4324]{A.~Nayerhoda}
\affiliation{Institute of Nuclear Physics Polish Academy of Sciences, PL-31342 IFJ-PAN, Krakow, Poland }
\author[0000-0003-1059-8731]{L.~Nellen}
\affiliation{Instituto de Ciencias Nucleares, Universidad Nacional Autónoma de Mexico, Ciudad de Mexico, Mexico }
\author[0000-0001-9428-7572]{M.~Newbold}
\affiliation{Department of Physics and Astronomy, University of Utah, Salt Lake City, UT, USA }
\author[0000-0001-7099-108X]{R.~Noriega-Papaqui}
\affiliation{Universidad Autónoma del Estado de Hidalgo, Pachuca, Mexico }
\author{A.~Peisker}
\affiliation{Department of Physics and Astronomy, Michigan State University, East Lansing, MI, USA }
\author[0000-0002-8774-8147]{Y.~Pérez Araujo $^{c}$}
\affiliation{Instituto de Astronomía, Universidad Nacional Autónoma de México, Ciudad de Mexico, Mexico, \href{mailto:nifraija@astro.unam.mx}{$^{a}$nifraija@astro.unam.mx}, \href{mailto:nimagda@astro.unam.mx}{$^{b}$magda@astro.unam.mx}, \href{mailto:yfperez@astro.unam.mx}{$^{c}$yfperez@astro.unam.mx}}
\author[0000-0001-5998-4938]{E.G.~Pérez-Pérez}
\affiliation{Universidad Politecnica de Pachuca, Pachuca, Hgo, Mexico }
\author{Z.~Ren}
\affiliation{Dept of Physics and Astronomy, University of New Mexico, Albuquerque, NM, USA}
\author[0000-0002-6524-9769]{C.D.~Rho}
\affiliation{Department of Physics \& Astronomy, University of Rochester, Rochester, NY , USA }
\author[0000-0003-1327-0838]{D.~Rosa-González}
\affiliation{Instituto Nacional de Astrof\'isica, Óptica y Electrónica, Puebla, Mexico }
\author{M.~Rosenberg}
\affiliation{Department of Physics, Pennsylvania State University, University Park, PA, USA }
\author[0000-0001-5079-5559]{J. R. Sacahui}
\affiliation{Instituto de Investigación en Ciencias Físicas and Matemáticas USAC, Ciudad Universitaria, Zona 12, Guatemala}
\author{H.~Salazar}
\affiliation{Facultad de Ciencias Físico Matemáticas, Benemérita Universidad Autónoma de Puebla, Puebla, Mexico }
\author{F.~Salesa Greus}
\affiliation{Institute of Nuclear Physics Polish Academy of Sciences, PL-31342 IFJ-PAN, Krakow, Poland}
\affiliation{Instituto de Física Corpuscular, CSIC, Universitat de Valéncia,E-46980, Paterna, Valencia, Spain}
\author[0000-0001-6079-2722]{A.~Sandoval}
\affiliation{Instituto de Física, Universidad Nacional Autónoma de México, Ciudad de Mexico, Mexico }
\author{J.~Serna-Franco}
\affiliation{Instituto de Física, Universidad Nacional Autónoma de México, Ciudad de Mexico, Mexico}
\author{A.J.~Smith}
\affiliation{Department of Physics, University of Maryland, College Park, MD, USA }
\author[0000-0002-1492-0380]{R.W.~Springer}
\affiliation{Department of Physics and Astronomy, University of Utah, Salt Lake City, UT, USA }
\author{P.~Surajbali}
\affiliation{Max-Planck Institute for Nuclear Physics, 69117 Heidelberg, Germany}
\author{O.~Tibolla}
\affiliation{Universidad Politecnica de Pachuca, Pachuca, Hgo, Mexico }
\author[0000-0001-9725-1479]{K.~Tollefson}
\affiliation{Department of Physics and Astronomy, Michigan State University, East Lansing, MI, USA }
\author[0000-0002-1689-3945]{I.~Torres}
\affiliation{Instituto Nacional de Astrofísica, Óptica y Electrónica, Puebla, Mexico }
\author{R.~Torres-Escobedo}
\affiliation{Departamento de Física, Centro Universitario de Ciencias Exactase Ingenierias, Universidad de Guadalajara, Guadalajara, Mexico }
\author{R.~Turner}
\affiliation{Department of Physics, Michigan Technological University, Houghton, MI, USA }
\author[0000-0001-6876-2800]{L.~Villaseñor}
\affiliation{Facultad de Ciencias Físico Matemáticas, Benemérita Universidad Autónoma de Puebla, Puebla, Mexico }
\author{X.~Wang}
\affiliation{Department of Physics, Michigan Technological University, Houghton, MI, USA }
\author[0000-0002-6623-0277]{E.~Willox}
\affiliation{Department of Physics, University of Maryland, College Park, MD, USA }
\author{A.~Zepeda}
\affiliation{Physics Department, Centro de Investigacion y de Estudios Avanzados del IPN, Mexico City, DF, Mexico }
\author[0000-0003-0513-3841]{H.~Zhou}
\affiliation{Physics Division, Los Alamos National Laboratory, Los Alamos, NM, USA }
\author{THE HAWC COLLABORATION}

\begin{abstract}
Many gamma-ray bursts (GRBs) have been observed from radio wavelengths, and a few at very-high energies (VHEs, $> $ 100GeV). 
The HAWC gamma-ray observatory is well suited to study transient phenomena at VHEs due to its large field of view and duty cycle. These features allow for searches of VHE emission and can probe different model assumptions of duration and spectra. In this paper, we use data collected by HAWC between December 2014 and May 2020 to search for emission in the energy range from 80 to 800 GeV coming from a sample  47 short GRBs that triggered the {\it Fermi}, {\it Swift} and {\it Konus} satellites during this period. This analysis is optimized to search for delayed and extended VHE emission within the first 20 s of each burst. We find no evidence of VHE emission, either simultaneous or delayed, with respect to the prompt emission. Upper limits (90\% confidence level) derived on the GRB fluence are used to constrain the synchrotron self-Compton forward-shock model. Constraints for the interstellar density as low as $10^{-2}\ \mathrm{cm}^{-3}$ are obtained when assuming z=0.3 for bursts with the highest keV-fluences such as GRB 170206A and GRB 181222841. Such a low density makes observing VHE emission mainly from the fast cooling regime challenging.
\end{abstract}


\section{Introduction} \label{sec:intro}
Gamma-ray bursts (GRBs), the most luminous events in the universe, are characterized by non-repeating flashes with a wide range of spectral and temporal features. They are classified in two groups based on the hardness ratio\footnote{
Defined as the ratio of the flux in two separated energy bands} \citep{2000PASJ...52..759Q,2001A&A...369..537Q} and the duration\footnote{Defined as $T_{90}$, the time during which the cumulative number of detected counts above background increases from 5\% to 95\%.}  of their prompt emission \citep{1993ApJ...413L.101K}. Bursts with duration longer and shorter than 2 s are defined as long and short bursts, respectively. Short GRBs typically present a higher hardness ratio with respect to the long class. The two classes are associated to different types of progenitor. Long GRBs are directly connected to the final phases of the life of massive stars \citep{1993ApJ...405..273W, 1999ApJ...524..262M} and short GRBs are related to compact-object mergers \citep[black hole - neutron star or neutron star - neutron star;][]{1989Natur.340..126E, 1992ApJ...395L..83N}, and detection of gravitational wave events \citep{2017ApJ...848L..12A}. 

GRB emission exhibits early and late phases. The early phase, so-called prompt emission, is usually detected in the energy range of a few keV to a few MeV with a range of light curve morphologies and variabilities. The late phase, referred to as afterglow, observed at multi-many wavelengths, from radio to $\gamma$-rays, appears after and lasts longer than the prompt emission. 

The theory most usually invoked to describe GRBs is the fireball model \citep{1986ApJ...308L..43P,1986ApJ...308L..47G,1978MNRAS.183..359C}.  It explains the prompt emission as dissipation of kinetic energy in internal shocks and the afterglow by the collision of the expanding blast wave with the external medium \citep[see ][ for a complete review]{2004RvMP...76.1143P}. Synchrotron radiation is considered as the cooling mechanism for both prompt and afterglow emissions. However, for 
prompt emission some discrepancies between observations and theory remain.  High ($> 10$ GeV) and very-high-energy (VHE, $> 100$ GeV) emission are described (for high emission) and predicted (for VHE emission) by either leptonic or hadronic models.

For leptonic models,  VHE emission is predicted to appear 
delayed with respect to the prompt emission as a result of synchrotron self-Compton (SSC) emission in external forward shocks  \cite{2001ApJ...548..787S,2001ApJ...546L..33W, 2001ApJ...559..110Z, 2012ApJ...755...12V, 2019ApJ...883..162F, 2019ApJ...879L..26F, 2021ApJ...918...12F} and external reverse shocks \citep{2003ApJ...598L..11G,2001ApJ...546L..33W,2001ApJ...556.1010W}. The delay is understood as the time for the shock to approach the deceleration radius \citep{1994MNRAS.269L..41M} and as a consequence of $\gamma\gamma$-opacity effects \citep[e.g., see][] {2006ApJ...650.1004B}. 

For hadronic models,  $\gamma$-ray emission  has been explained through photo-hadronic interactions between high-energy hadrons accelerated in the jet  and internal synchrotron photons \citep{2009ApJ...705L.191A, 2000ApJ...537..255D},   inelastic proton-neutron collisions  \citep{2000ApJ...541L...5M}   and  interactions of high-energy neutrons with photons from the jet \citep{2004A&A...418L...5D, 2004ApJ...604L..85A}. Even though GRBs are among the most plausible candidates to accelerate cosmic rays up to ultra-high energies \citep[$\gtrsim 10^{18}$ eV; ][]{1995PhRvL..75..386W, 1995ApJ...453..883V} and thus, potential candidates  for  neutrino  detection, the IceCube collaboration reported  no coincidences between neutrinos and GRBs after analyzing years of data \citep{2022arXiv220511410A,2012Natur.484..351A, 2016ApJ...824..115A, 2015ApJ...805L...5A}. Therefore, we conclude
that the number of hadrons are low enough that  hadronic interactions are inefficient at producing detectable $\gamma$-ray signals in GRBs and we accordingly exclude hadronic models from the interpretation of our results. 

High-energy emission has been reported for more than 186 GRBs\footnote{https://heasarc.gsfc.nasa.gov/W3Browse/fermi/fermilgrb.html} (from over a thousand bursts observed at keV energies) by the Large Area Telescope instrument on board the {\it Fermi} satellite \citep[{\it Fermi}-LAT;][]{2019ApJ...878...52A, 2009ApJ...697.1071A}. In some cases such high-energy emission is consistent with an extrapolation of the prompt component observed at keV - MeV \citep{2013ApJS..209...11A} and in others it also exhibits a spectral component from hundreds of MeV to a few GeV with different evolution with respect to the prompt emission \citep[eg.;][]{2010ApJ...716.1178A,2009ApJ...706L.138A}. Furthermore, {\it Fermi}-LAT observes an offset with respect to the beginning of the keV prompt phase and a long MeV-GeV emission lasting from hundreds to thousands of seconds after the trigger \citep[e.g.;][]{2009Sci...323.1688A,2009ApJ...706L.138A,2010MNRAS.409..226K}. Because of its temporal features, this MeV-GeV component seems strongly correlated to the X-ray afterglow \citep{2016ApJ...822...68A} in agreement with synchrotron external-shock emission. However, photons with energies higher than 10 GeV were observed in several long GRBs \citep[e.g. GRB 130427A, GRB 090926A, GRB 110731A;] []{2014Sci...343...42A,2011ApJ...729..114A,2013ApJ...763...71A} and in a short GRB \citep[GRB 090510;][]{2010ApJ...716.1178A} and may be evidence for inverse Compton (IC) scattering. If so, extrapolations of this IC component could lead to VHE photons that may be observable depending on the burst redshift and brightness. 

There have been several attempts to detect GRB counterparts at very high energies in the last 20 years. However, most observations yielded upper limits on the VHE flux  \citep[][]{2007ApJ...667..358A, 2014MNRAS.437.3103A, 2007ApJ...666..361A, 2009A&A...495..505A, 2009ApJ...690.1068A, 2014A&A...565A..16H, 2011ApJ...743...62A, 2015ApJ...800...78A, 2017ApJ...843...88A, 2017ApJ...842...31B, 2018arXiv180301266A}. The first claim of possible VHE emission was associated with GRB 970217A (with marginal significance) reported by the extensive air shower array Milagrito \citep{2000ApJ...533L.119A}. In the last four years, the striking detections of GRB 180720B, GRB 190829A and GRB 190114C above energies of 100, 200 and 300 GeV  by the H.E.S.S. and MAGIC observatories \citep{2019Natur.575..464A, 2019Natur.575..459A}, respectively,  strengthens the expectations of emission at VHE energies.   H.E.S.S observed GRB 180720B ten hours after the end of the prompt emission phase, which lasted $48.9\pm0.4\,{\rm s}$.  MAGIC observed GRB 190114C for $\sim$ 40 minutes, much longer than the prompt emission  ($T_{90}\simeq 25\,{\rm s}$). The H.E.S.S. telescopes reported the detection of VHE gamma-rays above $\geq$ 200 GeV with statistical significance of 21.7$\sigma$ during first night in the direction of GRB 190829A \citep{2021arXiv210602510H}.

The High Altitude Water Cherenkov (HAWC) gamma-ray observatory is an extensive air shower array located at Sierra Negra, in the state of Puebla, Mexico. Thanks to its wide field of view ($\sim$ 2 sr) and its continuous operation ($\sim 95 \%$ duty cycle), HAWC constantly searches for VHE emission from bursts detected by satellite instruments in its field of view. Since it does not need to be re-pointed to the GRB position, data before and after the GRB trigger are available,
making it possible to probe different model predictions for duration and spectra of VHE emission. Previously, VHE emission simultaneous to the prompt keV phase has been searched by HAWC \citep{2017ApJ...843...88A} for bursts between December 2014 and June 2016. Positive detection was not claimed and upper limits for prompt VHE emission were placed. 
In this paper we take into account that if GeV and VHE emissions come from interactions with the interstellar medium then they should show the same temporal behavior. This has to be delayed and last longer than the keV prompt phase.  Thus, in this paper we search for VHE emission with a strategy optimized for delayed and extended \footnote{Meaning that last longer that the duration of the prompt emission.} signals longer that the duration of the prompt emission.

We focus our analysis on short bursts for three reasons. First, VHE observations are challenging for long GRBs, usually observed at $z>1$ \citep{2005ApJ...634..501B,2006A&A...447..897J}, because of the spectral attenuation at energies above $\sim$100 GeV by the extragalactic background light (EBL) absorption \citep[as described by][]{2011MNRAS.410.2556D,2012MNRAS.422.3189G,2017A&A...603A..34F}. However, for short GRBs the average redshift is $\sim$0.48 \citep[][]{2014ARA&A..52...43B}. Second, HAWC is more sensitive to short GRBs rather than long GRBs \citep{2014NIMPA.742..276T} because the shorter duration of the search window results in a lower number of background events. Interestingly, two of the brightest short GRBs detected by {\it Fermi}-GBM, GRB 170206A and GRB 181222841 are within our set of short GRBs to study. Upper limits to the prompt emission of GRB 170206A were presented by \cite{2017ApJ...843...88A}. Third, the explicit expressions for light curves of the afterglow emission in the SSC model described in Section~\ref{sec:external_shock} 
are  developed under the assumption of a homogeneous medium which is unlikely for long GRBs \citep{2015PhR...561....1K} but probably the case for short bursts. Hereafter, we refer to short GRBs as sGRBs.

The paper is organized as follows. In Section~\ref{sec:data_analysis}, we present details of the analysis and results of the VHE emission search. In Section~\ref{sec:external_shock}, we present the basis of the SSC forward shock model used for the interpretation of our results. In Section~\ref{sec:upper_limits}, we discuss the HAWC flux upper limits in the SSC framework for the most relevant GRBs. Finally, conclusions are reported in Section~\ref{sec:conclusions}.  

\section{Search for VHE emission with HAWC} \label{sec:data_analysis}

As a first step, we look for VHE emission from our sample of 47 short GRB, listed in Table ~\ref{tbl:positions}, detected by the {\it Swift} and/or the {\it Fermi} satellite in HAWC's field of view between December 1st, 2014 and May 14, 2020. Table~\ref{tbl:positions} summarizes the GRB information of all bursts analyzed in this paper.  Since the positional uncertainty for bursts detected by the Gamma-ray Burst Monitor on board of {\it Fermi} ({\it Fermi}-GBM) is larger than the HAWC point spread function, the approach described in \cite{2017ApJ...843...88A} is adopted.  The remaining bursts are detected by the {\it Fermi}-LAT and {\it Swift} Burst Area Telescope (BAT) which provide localizations smaller than the HAWC point-spread function.

It is worth noting three VHE gamma-ray observations. The long burst GRB 190114C showed emission starting $\sim$ 60 s after the trigger burst and lasted $\sim$ 40 minutes \citep{2019Natur.575..464A}. In the case of GRB 180720B and GRB 190829A, the VHE emission was detected by the MAGIC and H.E.S.S. observatories ten and four hours after the trigger, respectively. In all three cases, the detection was interpreted in the SSC forward-shock scenario and did not always come during the prompt emission \citep{2019Natur.575..459A}. Inspired by long bursts, we implement a modified strategy to the search presented by \cite{2017ApJ...843...88A} for prompt emission. We keep the methodology of searching on one-degree multiple circles, each of them offset by 0.3 degree in right ascension or declination until covering the uncertainty of the burst position. The background is estimated by scaling the all-sky event rate for a one-degree circle. When this is not possible, the mean of at least ten independent off-observations \footnote{the search circle is offset in right ascension and time covering the same zenith angles as the search on the GRB position but at different time.} is taken. 

The search is optimized for delayed "afterglow" emission from short GRBs, considering the expected light curves (see Section \ref{sec:external_shock}). Instead of searching for VHE prompt emission starting on the trigger burst and with a duration of $T_{\rm 90}$, $3\times T_{\rm 90}$ and $10\times T_{\rm 90}$,  we search for VHE afterglow emission,  that could come late and last from one to 20 seconds. 
We use  ten consecutive time windows, each with a duration of 2 s. The sensitivity of this search was tested with Monte Carlo simulation by considering different starting times, duration values and intensities of the bursts.
We find an improvement of a factor of 4 in the sensitivity of the search, when looking for afterglow emission, compared to \cite{2017ApJ...843...88A}, assuming a type II error probability of 10\%.

Only relevant details of the adopted analysis are given here; for a full description see \cite{2017ApJ...843...88A}. When a minimum number of photomultiplier tubes (PMTs) trigger within a given time window, a HAWC event is recorded. The events are classified in nine different size bins (from 1 to 9) defined by the fraction of available PMTs triggered by the event \cite[see][for details]{2017ApJ...843...39A}. The Gamma/Hadron separation criteria summarized in \cite{2017ApJ...843...39A} are applied from bin 1 to bin 9. The search for signal is performed in ten consecutive time windows of 2 s to preserve a minimum number of counts (for instance, the background rate is 0.54 events per second for GRB 150423A) \citep[see ][]{2017ApJ...843...88A}. The significance is calculated applying the search in one-degree circles as described in \cite{2017ApJ...843...88A} and assuming a Poisson distribution of the background rate. It is expressed in terms of standard deviations of Gaussian distribution equivalent to the corresponding \textit{p}-value. When no events are observed, the significance is not calculated. This is the case for bursts GRB 150811849, GRB 160714A, GRB 170826369 and GRB 191031891: none of these are within 20 degrees of zenith, where HAWC is most sensitive.

\begin{figure}[ht!]
\epsscale{0.8}
\plotone{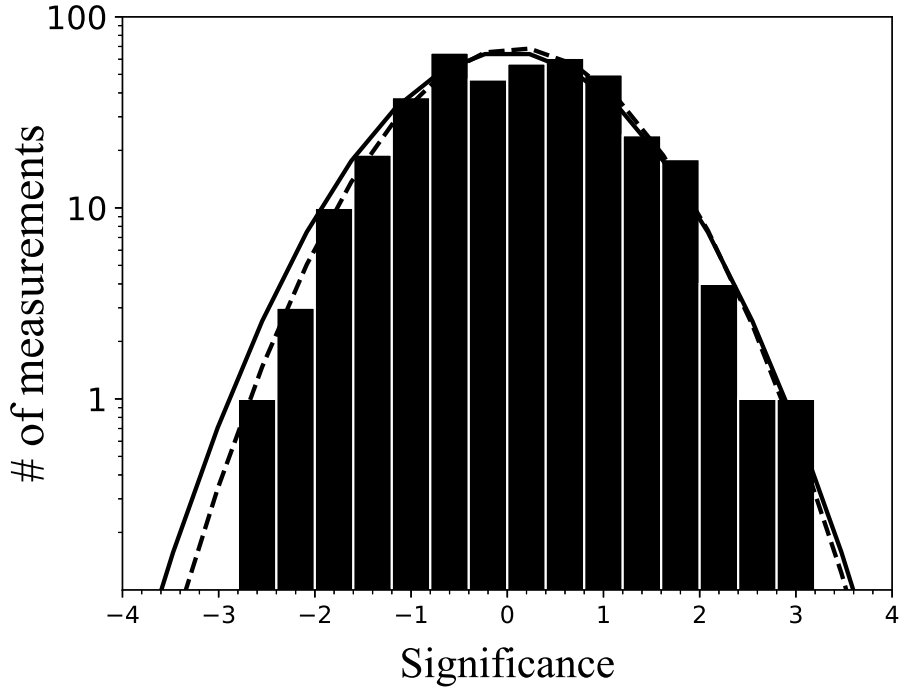}

\caption{Distribution of the significances after trial correction extracted for each GRB of the total sample on the ten consecutive time windows. It is consistent with a normal distribution.  The solid line represents a normal distribution with mean of zero and standard deviation of 1. The normal distribution that fits the significance distribution has a mean of $0.09 \pm 0.05$ and a standard deviation of $0.95 \pm 0.03$, shown as dashed line.\\  
\label{fig:significance}}
\end{figure}

The distribution of the significances extracted for the analyzed sample of GRBs in all time windows is shown in Figure ~\ref{fig:significance} and individual values for the first time window are stated in Table ~\ref{tbl:positions}  and only those time windows with significance greater than 2$\sigma$ are given in Table ~\ref{tbl:HighSignificance}. In both cases, the position of the corresponding one-degree circle is given. To complete the results in \cite{2017ApJ...843...88A}, HAWC upper limits for an spectral index of -0.5 in the energy range of 80 to 800 GeV are also given. The highest significance post trial of 3.15$\sigma$ (4.78 pre trials) is obtained for GRB 200514380 in the time window from 14 to 16 s after the trigger time. The significance during the $T_{90}$ of each GRB is also calculated, finding a maximum significance of 2.39$\sigma$ for GRB170604603.

The given significance is after trial correction. Considering the 399 time windows, we expect $\sim$0.52 fluctuations above 3.0 $\sigma$, meaning we find no evidence for a positive detection at either early or late times. Then, event upper limits are derived as upper edge of frequentist confidence intervals of 90\% and converted to flux limits as described in \cite{2017ApJ...843...88A} under the spectral assumptions described in Section~\ref{sec:external_shock} to constrain the parameters of the theoretical model. Flux upper limits for the first time window of each burst is given in Table ~\ref{tbl:positions}.

\section{SSC Forward-shock Model}
\label{sec:external_shock}
The dynamics of the afterglow emission is generally modeled as a spherical ultra-relativistic and adiabatic shell propagating into a homogeneous density \citep{1998ApJ...497L..17S, 1995ApJ...455L.143S, 2000ApJ...532..286K,1999A&AS..138..537S}. In particular, \citet{1998ApJ...497L..17S} describe the spectrum and light-curves for the synchrotron radiation, and the inverse Compton model is treated in detail by \cite{2000ApJ...544L..17P} and later extended for the slow-cooling regime by \cite{2000ApJ...532..286K}. Therefore, here we show explicitly the spectral breaks, the maximum flux and the light curves for the fast and slow-cooling regimes as follows.

Accelerated electrons can up-scatter photons from low to high energies proportionally to the square of the electron Lorentz factor, $\gamma^2_e$. Considering the electron Lorentz factors, the synchrotron spectral breaks and the maximum flux given by \citet{1998ApJ...497L..17S}, the characteristic ($E^{\rm ssc}_{\rm \gamma,m}$) and cooling ($E^{\rm ssc}_{\rm \gamma,c}$) break energies in the SSC emission can be written as 
{\small
\bary\label{sscf}
E^{\rm ssc}_{\rm \gamma,m}&\simeq& A_{\rm \gamma,m}\,  (1+z)^{\frac54}\,\epsilon_e^{4}\,\varepsilon_{B}^{\frac12}\,n^{-\frac14}\,E^{\frac34}\,t^{-\frac94}\cr
E^{\rm ssc}_{\rm \gamma,c}&\simeq& A_{\rm \gamma,c}\, (1+z)^{-\frac34}\,(1+Y)^{-4}\,\varepsilon_{B}^{-\frac72}\,n^{-\frac94}\,E^{-\frac54}\,t^{-\frac14}\,,
\eary
}
and the maximum flux is given as,
{\small
\bary
F^{\rm ssc}_{\rm \gamma,max}&\simeq& F_{\rm \gamma,max}\,(1+z)^{\frac34}\,\varepsilon_{B}^{\frac12}\,n^{\frac54}\,D_{\rm z}^{-2}\,E^{\frac54}\, t^{1/4}\,.
\eary
}
The coefficients of the break energies ($A_{\rm \gamma,m}$ and $A_{\rm \gamma,c}$) and the maximum flux ($F_{\rm \gamma,max}$) are given explicitly in the Appendix. Here, $E$ is the isotropic kinetic energy of the  blast  wave, $n$ is the density of  the  surrounding  medium, $z$ is the redshift, $p$ is the electron spectral index, $D_{\rm z}$ is the luminosity distance from the burst to Earth, $Y$ is the Compton parameter \citep{2001ApJ...548..787S} and, $\epsilon_{e}$ and $\epsilon_{B}$ are the microphysical parameters related to 
the total energy given to accelerate electrons and to amplify the magnetic field, respectively, with the constraint of  $\epsilon_{e} + \epsilon_{B} < 1$.

The light curves in the fast (f) and slow (s) cooling regime\footnote{The difference between the two regimes is associated with radiative timescale of this emission \citep[e.g. see,][]{2001ApJ...548..787S}} are given by,

{\small
\begin{eqnarray}
\label{fcssc_t}
F_{\nu}= \cases{ 
A_{\rm f1}\,t^{\frac{1}{3}}\, \left(E_\gamma\right)^{\frac13},\hspace{1.3cm} E_\gamma<E_{\rm \gamma,c}, \cr
A_{\rm f2}\,t^{\frac{1}{8}}\, \left(E_\gamma\right)^{-\frac12},\hspace{0.7cm}\,\,\,\, E_{\rm \gamma,c}<E_\gamma<E_{\rm \gamma,m}, \cr
A_{\rm f3}\,t^{-\frac{9p-10}{8}}\,\left(E_\gamma\right)^{-\frac{p}{2}},\,\,\,\,E_{\rm \gamma,m}<E_\gamma<E_{\rm \gamma,max}\,, \cr
}
\end{eqnarray}
}
and
{\small
\begin{eqnarray}
\label{scssc_t}
F_{\nu}=\cases{
A_{\rm s1}t\left(E_\gamma\right)^{\frac13},\hspace{2.14cm} E_\gamma<E_{\rm \gamma,m},\cr
A_{\rm s2}t^{-\frac{9p-11}{8}}\left(E_\gamma\right)^{-\frac{p-1}{2}},\hspace{0.5cm}E_{\rm \gamma,m}<E_\gamma<E_{\rm \gamma,c},\,\,\,\,\,\cr
A_{\rm s3}\,t^{-\frac{9p-10}{8} + \frac{p-2}{4-p}}\,\left(E_\gamma\right)^{-\frac{p}{2}},\,E_{\rm \gamma,c}<E_\gamma<E_{\rm \gamma,max}\,, \cr
}
\end{eqnarray}
}
respectively. The coefficients $A_{\rm f1}$, $A_{\rm f2}$, $A_{\rm f3}$, $A_{\rm s1}$, $A_{\rm s2}$ and $A_{\rm s3}$, are given explicitly in the Appendix.

The Klein-Nishina (KN) correction in the spectrum must be considered 
at very high energies where SSC emission decreases drastically. The break energy in the KN regime is given by 
{
\small
\be
E^{\rm KN}_{\rm \gamma,c}\simeq A^{\rm KN}_{\rm \gamma,c}(1+z)^{-1}\, (1+Y)^{-1} \,\varepsilon_{B,f}^{-1}\,n^{-\frac23}\,\Gamma^\frac23\,E^{-\frac13}\,t^{-1/4}\,,
\ee
}
where $\Gamma$ is the bulk Lorentz factor given by
\begin{equation}\label{Gamma}
\Gamma=\left(\frac{3}{32\,\pi\,m_p}\right)^{1/8}\,(1+z)^{3/8}\,n^{-1/8}\,E^{1/8}\,t^{-3/8}\,.
\end{equation}
 
The SSC light curves for fast- and slow- cooling regimes are derived assuming that the scattering occurs in a non-relativistic regime (below the KN limit). The term $m_{\rm p}$ is the proton mass. Specifically, we require the KN energy break ($E^{KN}_{\rm \gamma,c}$) to be higher than 1 TeV.

We do not take into account the intrinsic attenuation by $e^\pm$ pair production because the outflow is in the deceleration phase.  We introduce the attenuation produced by EBL absorption in accordance with the model presented in \cite{2012MNRAS.422.3189G}. We impose the restriction of $\epsilon_{e}>\epsilon_{B}$ to assure that the inverse Compton mechanism is efficient.

Theoretical light curves are calculated from Equations \ref{fcssc_t} and \ref{scssc_t} varying the parameters $\epsilon_{\rm B}$, $\epsilon_{\rm e}$ and $n$ within the ranges of $[10^{-5}, 10^{0})$, $[10^{-2}, 10^{0})$ \citep{2014ApJ...785...29S} and $[10^{-4}, 10^0]\,{\rm cm^{-3}}$ \citep{2006ApJ...650..261S, 2014ARA&A..52...43B}, respectively. 
The kinetic energy is obtained from the isotropic energy assuming a kinetic energy efficiency of 20\%  \citep{2015PhR...561....1K}. The isotropic energy in gamma-rays is given by the expression,\\
\be \label{eq:Eiso}
E_{\rm iso}= \frac{4\pi D_{\rm z}^{2} F_{\gamma}}{(1+z)}
\ee 
where $D_{z}$ is the luminosity distance, $F_{\gamma}$ is the fluence in $\gamma$-rays detected by Fermi-GBM  and $z$ is the redshift. We assume  for the cosmological constants a spatially flat universe $\Lambda$CDM model with  $H_0=67.4\,{\rm km\,s^{-1}\,Mpc^{-1}}$, $\Omega_{\rm M}=0.315$ and  $\Omega_\Lambda=0.714$ \citep{2016A&A...594A..13P}. 

\begin{figure}[ht!]
\epsscale{0.8}
\plotone{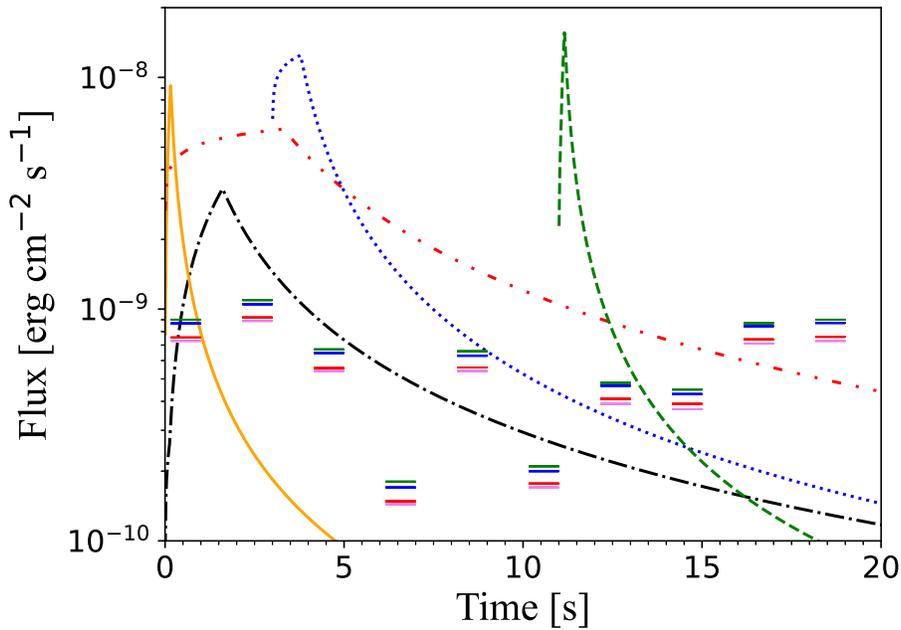}
\caption{For illustrative purposes the flux as a function of time predicted by the SSC model as described in Section~\ref{sec:external_shock} is shown for arbitrary values of microphysical parameters for GRB 170206A. Red(dashdotted), blue(dotted) and green(dashed) lines show the theoretical light curves in the fast cooling regime assuming different combination of microphysical parameters ([$\varepsilon_{B}$=$1.4\times10^{-2}$, $\varepsilon_{e}$=$2.6\times10^{-2}$],[$\varepsilon_{B}$=$6.5\times10^{-3}$, $\varepsilon_{e}$=$1.3\times10^{-2}$] and [$\varepsilon_{B}$=$5.7\times10^{-4}$, $\varepsilon_{e}$=$7.1\times10^{-3}$], respectively) and different start times ($t_{\rm start}=0$ sec, $t_{\rm start}=3$ sec and $t_{\rm start}=11$ sec, respectively). Slow-cooling regime light curves, plotted in orange(solid) and black(dashdot) are derived assuming [$\varepsilon_{B}$=$1.9\times10^{-4}$, $\varepsilon_{e}$=$8.0\times10^{-3}$] and [$\varepsilon_{B}$=$7.8\times10^{-6}$, $\varepsilon_{e}$=$4.5\times10^{-2}$], respectively. For all the cases we assume a redshift of $z=0.3$, n = 1 ${\rm cm^{-3}}$ and the isotropic energy of $E_{\rm iso}=3.6\times10^{51}\,{\rm erg}$. Symbols represent flux upper limits at observation energy of 500 GeV derived from HAWC upper limits for fluences in the energy range of 80 to 800 GeV. The color of symbols correspond to different spectral indexes assumptions, red for 0.7, green for 1.5, blue for 1.7 and violet for 2.2. 
\label{fig:th_lc}}
\end{figure}

Figure ~\ref{fig:th_lc} shows theoretical light curves for different start times and microphysical parameter values. The observed emission can be as short as the green light curve or long as the red light curve. Moreover, the peak of the emission can appear along the 20 s search period. The analysis presented here takes into account all possible light curve profiles within the model. It is more restrictive to constrain the spectral time evolution than using a unique time window that only contains information on the total fluence. Therefore, for each time window, HAWC upper limits for fluences in the energy range of 80 to 800 GeV are calculated for the spectral indexes corresponding to the theoretical light curves. Then, fluxes at observation energy of 500 GeV are obtained (symbols in Figure ~\ref{fig:th_lc}) and compared with fluxes expected of the theoretical light curves (lines in Figure ~\ref{fig:th_lc}) at the midpoint of the time interval of each time window and the same observation energy as shown in Figure ~\ref{fig:th_lc}. This allows us to constrain physical parameter values such as the interstellar medium (ISM) density or microphysical parameters. For a more detailed analysis, see \citep{2021arXiv210803333P}. 

\section{Results and Discussion}
\label{sec:upper_limits}

For theoretical fluences, we consider a typical value  of the electron spectral index for forward shocks $p = 2.4$, \citep[see, e.g.][]{2015PhR...561....1K}. Also, since there are no measurements of redshifts for the GRBs considered in this analysis, we assume a value of $z=0.3$ to derive the theoretical fluences and the HAWC upper limits. However, the dependency of HAWC upper limits with the assumed redshift is discussed later. The assumption is motivated by the observed distribution of redshifts \citep[see][for a review]{2014ARA&A..52...43B} and the previous work of \cite{2017ApJ...843...88A} where it is shown that $90\%$ of the photons expected from the source would have an energy between 80 and 800 GeV as an effect, mainly, of the EBL. For z = 0.3 a $50\%$ attenuation on the spectra is expected at energies of $\sim 400$ GeV while for a z = 0.5 and z = 1, a same attenuation is expected at energies $\lesssim 150$ GeV and $\lesssim 100$ GeV \cite{2012MNRAS.422.3189G, 2017A&A...603A..34F}, correspondingly. Then a choice of z = 0.5 will give results similar to the ones for z = 1, where the EBL would absorb most of the emission detectable by HAWC. The chosen value of redshift is a good compromise 
between EBL, the mean redshift observed for short bursts and the loss of HAWC sensitivity (best above 1 TeV). 

\begin{figure}[ht!]
\epsscale{0.8}
\plotone{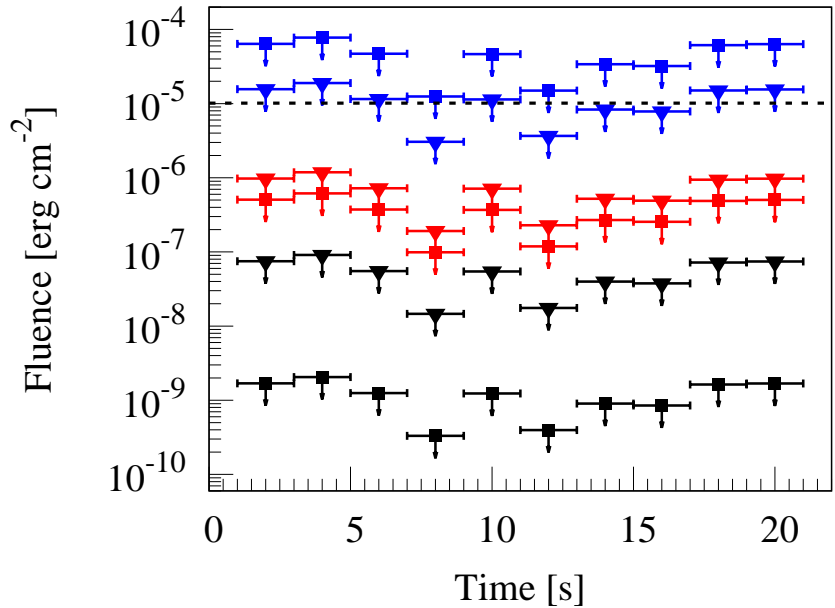}
\caption{HAWC upper limits of the fluence extracted in the energy range 80-800 GeV for the GRB 170206A for different spectral indexes and redshifts. The limits are calculated for spectral indexes of -0.7 (triangles), -2.2 (squares) and for redshifts of 0 (black), 0.3 (red) and 1 (blue). The dashed black line is the average flux measured by {\it Fermi}-GBM during the prompt emission   
\label{fig:170206_limits}}
\end{figure}

HAWC upper limits are calculated to be compared to the theoretical fluences considering the theoretical spectral indexes and the energy breaks determined by the parameter values. Figure~\ref{fig:170206_limits} shows HAWC upper limits of the fluence as function of redshift and the assumed intrinsic spectral index.  The variation of the upper limit values with time (for a same redshift) is due to signal fluctuations over the background level. As expected, the limits can vary up to four orders of magnitude when the redshift goes from 0 to 1. The smallest and largest variation of the HAWC upper limits are for a redshift of 0.3 and 0, respectively. This variation decreases as redshift increases. However, the most substantial attenuation of the spectra is over the observational energy range. This causes that for z = 1, the HAWC UL for the spectral index of -0.7 is below the ones for the spectral index of -2.2. As the redshift increases from z = 0.1 to z = 1, the optical depth, due to EBL, takes a value of 1 at energies of 900 GeV to 90 GeV, respectively.

\begin{figure}[ht!]
\epsscale{1.0}
\plotone{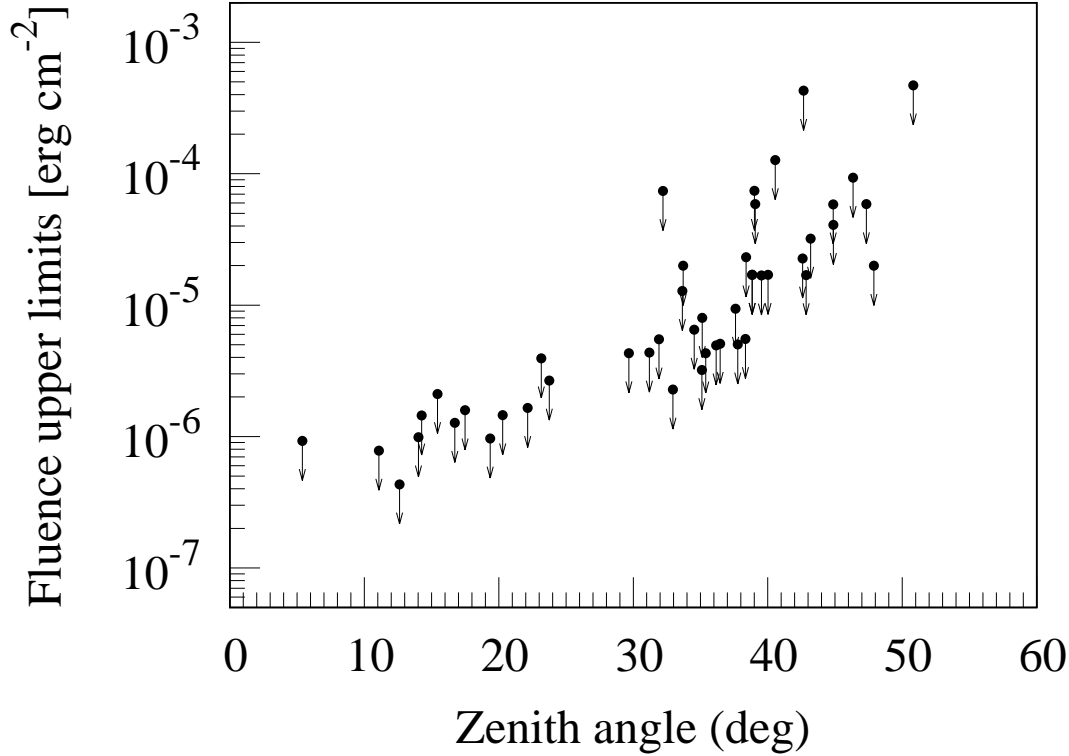}
\caption{Distribution of the fluence 90\%  upper limits extracted in the first time window (0-2 s) for the 43 short GRB positions with respect to the zenith angle of the source. All these limits are derived between 80--800 GeV assuming a redshift of 0.3 and a spectral index of -0.5. The detector sensitivity is best  up to $\sim20^{\circ}$. Bursts considered in this work and included in this Figure are reported in Table \ref{tbl:positions}.  
\label{fig:zenith}}
\end{figure}

To show the dependence of the upper limits with the zenith angle, the fluence upper limits for the first time window of each burst  as a function of the zenith angle for the spectral index of $-0.5$ and $z=0.3$ are shown in Figure ~\ref{fig:zenith}. The value used for the spectral index corresponds to the fast-cooling regime when the observation energy is between the cooling and characteristic break energies (second power law). For our analysis, the spectral index was varied to calculate the upper limits as mentioned above and explained later in Section ~\ref{sec:upper_limits}. As observed, there is a strong dependency on the zenith angle reflected in a variation of three orders of magnitude in the limits. This effect is expected and discussed in detail in \cite{2012APh....35..641A}. Therefore, here we have only considered bursts with zenith angles less than $21^\circ$, where the HAWC sensitivity is best.  These bursts are: GRB 141205A, GRB 150423A, GRB160406503, GRB 150710A, GRB 170206A, GRB 170709334, GRB 180103090, GRB 180617872, GRB 181222841 and  GRB 190905985. In the case of GBM-detected bursts, the tiling search method described by \cite{2017ApJ...843...88A} is used. Thus, we take the most conservative selection by choosing the fluence upper limit derived in the position inside the GBM error box with the highest significance. This selection contributes to the spread observed in Figure ~\ref{fig:zenith}.  

\begin{figure}[ht!]
\epsscale{1.0}
\plotone{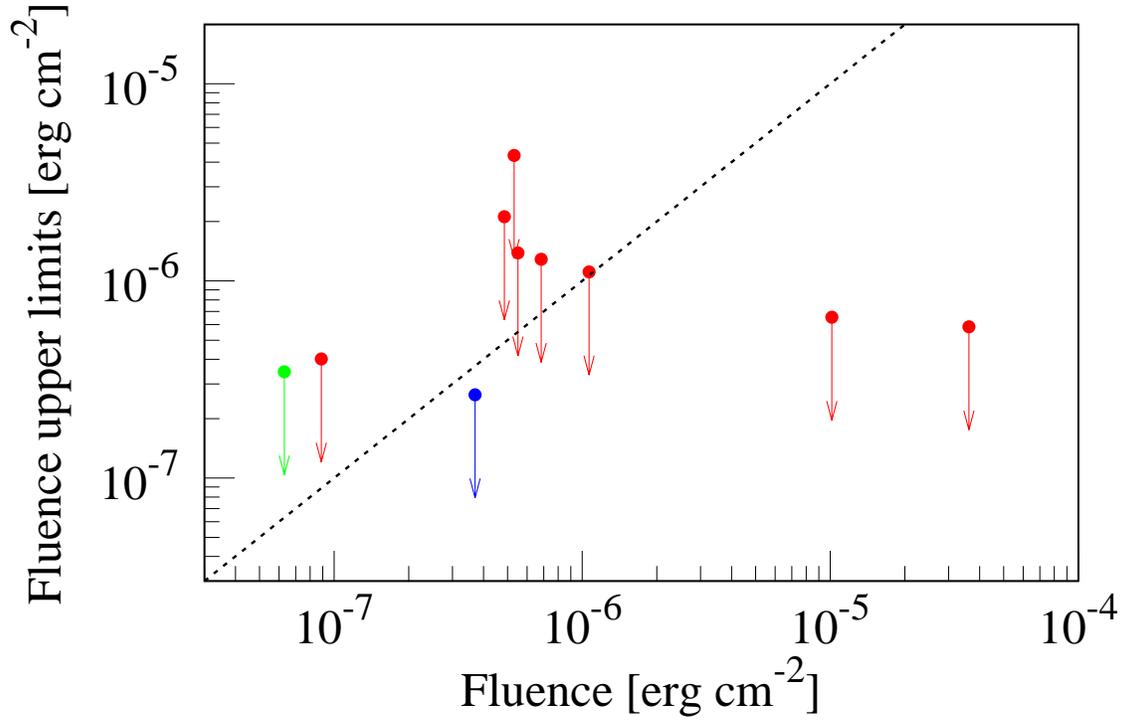}
\caption{The 90\% upper limits derived by HAWC for the sample of burst with zenith angle less than $21^\circ$ versus the fluence observed during the prompt emission are shown.  The corresponding energy ranges are: 15--150 keV ({\it Swift}-BAT, green), 10--1000 keV ({\it Fermi}-GBM, red) and 20--2000 keV ({\it Konus}/WIND, blue). 
HAWC UL are derived using a time window of T$_{90}$ (the same referred to the fluence provided by the satellite) and assuming a power-law spectrum with index -0.5 and $z=0.3$. The identity function is shown as black dotted line. We show, from left to right, GRB 150423A, GRB 180103090, GRB 200514380, GRB 150710A, GRB 180617872, GRB160406503, GRB 190905985, GRB 170709334, GRB 141205A, GRB 170206A and GRB 181222841.
\label{fig:limit_vs_prompt}}
\end{figure}

In Figure~\ref{fig:limit_vs_prompt}, we compare, for the bursts selected by zenith angle, the fluence upper limits derived by HAWC over a time interval of $T_{90}$ and a spectral index of $-0.5$ with the fluence measured\footnote{We use the data reported in the official {\it Fermi}-GBM catalog and GCNs to derive the prompt emission fluences \citep{2015GCN.17740....1U, 2017GCN.20443....1L, 2015GCN.18003....1G}.} by {\it Swift}-BAT, {\it Fermi}-GBM and {\it Konus}-WIND in the energy ranges of 15--150 keV, 10--1000 keV and 20--2000 keV, respectively. GRB 170206A  and GRB 181222841 clearly stand out because the HAWC ULs, assuming $z = 0.3$, are one order of magnitude lower than the fluence measured by {\it Fermi}-GBM. Since the redshift is assumed to be the same for all bursts and HAWC upper limits are of the same magnitude, the results, burst to burst, depend mainly on the kinetic energy available, i. e. on the fluence in the prompt phase measured by either  {\it Swift}-BAT, {\it Fermi}-GBM or {\it Konus}-WIND, as seen in Equation~\ref{eq:Eiso}. Thus, we select four representative bursts that sample the entire range of keV-fluences. These are GRB 170206A and GRB 181222841 (with the highest keV-fluence), GRB170709334 and, GRB 180103090 (with the lowest {\it Fermi}-GBM fluence). Table~\ref{tbl:fourgrbs} summarizes their relevant information for this analysis.

We consider different regimes and transitions between power-law spectra (see Eqs.~\ref{fcssc_t} and \ref{scssc_t}).  The emission begins in the fast-cooling regime and the spectral breaks are extremely high at the onset of the afterglow. As the spectrum evolves, several transitions take place, modifying the spectrum and the light curve. Three different scenarios are possible. If the transition between fast and slow cooling regime occurs before the time $t_c$ (the time when the spectral break $E^{\rm ssc}_{\rm \gamma,c}$ crosses below the considered energy range (in this case the HAWC energy range), when the transition occurs after the time $t_m$ (when $E^{\rm ssc}_{\rm \gamma,m}$ crosses below HAWC energy range) and, when the transition occurs after $t_c$ but before $t_m$. Therefore, for the analysis, three cases associated with these scenarios are defined: cases in the fast cooling regime during the first 20 s, cases in the transition regime where the flux evolves from fast to slow cooling regime after 2 s and before 20 s of the trigger time and, cases in the slow cooling regime from 2 to 20 s after the trigger time (since the afterglow starts in the fast cooling regime, the change to the slow cooling regime must happen in the first 2 s).

\begin{figure}[ht!]
\plotone{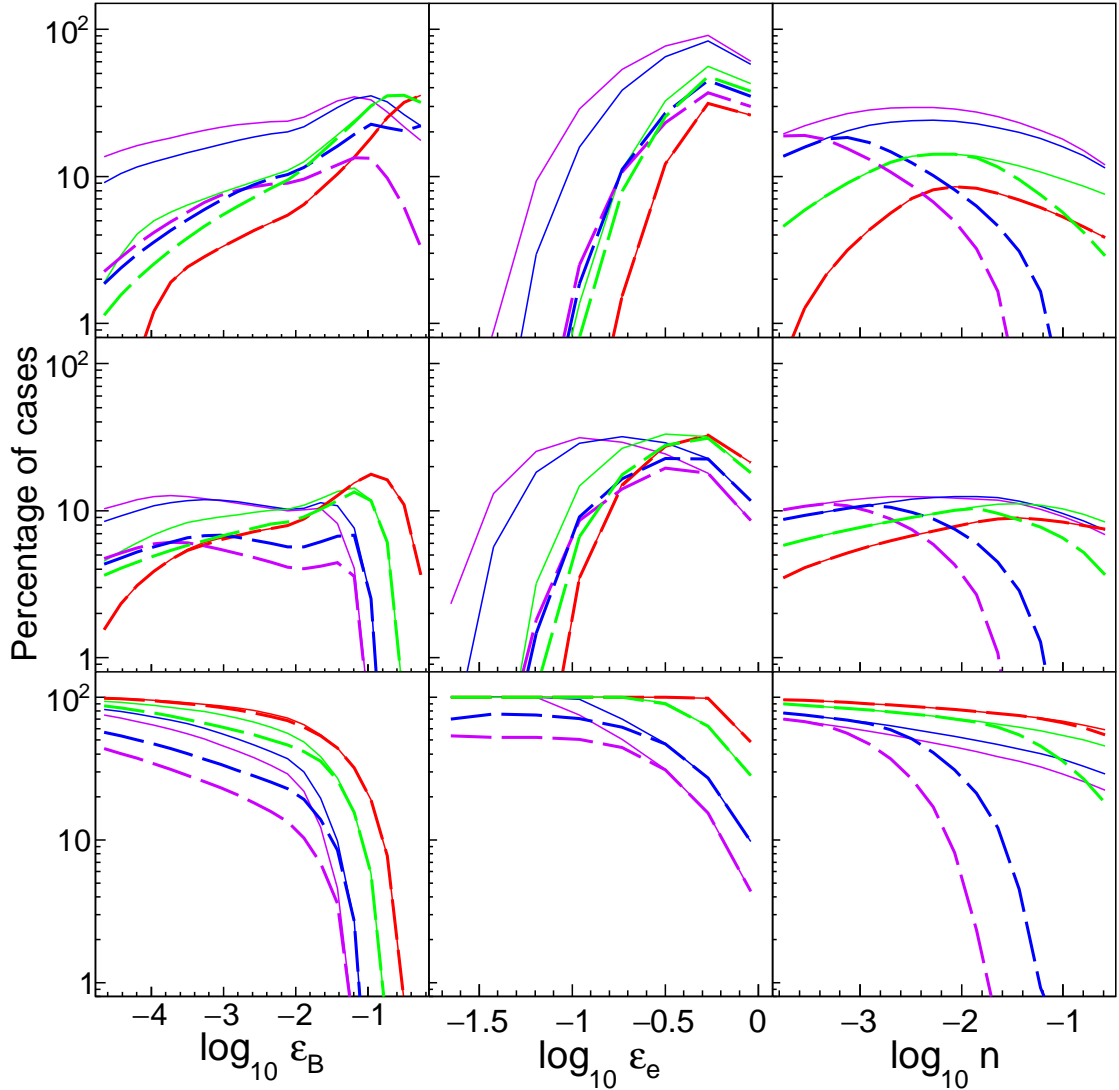}
\caption{Percentage of the parameter space in each cooling case as function of $\epsilon_{\rm B}$, $\epsilon_{\rm e}$ and ${\rm n}$ for fast (upper), slow (middle) and transition cases (lower) as defined in the text. Colors correspond to the keV-fluences of $ 3.62\times10^{-5} \rm erg\, cm^{-2}$ for GRB 181222A  (purple), $1.02\times10^{-5}\rm erg\, cm^{-2}$ for  GRB 170206A (blue), $6.83\times10^{-7}\rm erg\, cm^{-2}$ for GRB 170709A (green), and $8.88\times10^{-8}\rm erg\, cm^{-2}$ for GRB 180103A (red). Solid lines correspond to the cases considered for the given regime and keV-fluences while dashed lines correspond to the cases consistent with an hypothetical flux upper limit of $10^{-10} \rm erg\, \rm cm^{-2}$ at 500 GeV.
\label{fig:Histograms}}
\end{figure}

As mentioned before, our analysis considers two main ingredients, the HAWC upper limits on the flux and the theoretical light curves. It has been discussed how the flux upper limits depend on the zenith angle of observation, the assumed or measured redshift of the burst and the spectral indexes assumptions (in this analysis, set to match
the SSC forward shock model). In the case of the theoretical light curves, the set of parameters $n$, $\epsilon_{B}$, $\epsilon_{e}$ and the measured keV-fluence (through the apparent isotropic kinetic energy, see Eqn. \ref{eq:Eiso}) define the cooling case (see Eqn. \ref{fcssc_t} and \ref{scssc_t} and, appendix \ref{appendix}). In other words, the duration and intensity of the theoretical light curves are different from burst to burst because of their different keV-fluence. In order to understand, we consider the keV-fluences for the four bursts mentioned above. The percentages\textbf{\footnote{Calculated as the number of cases in a given cooling regime over the total number of cases, both for the range of parameter to be considered.}} of the parameter space in each cooling or transition case as function of the model parameters are shown as solid lines in Figure~\ref{fig:Histograms}.
 As observed, the slow cooling regime (middle panels) dominates the parameter space. For instance, for the highest (lowest) keV-fluence, shown as purple (red) solid line,  27  $\%$ (5  $\%$) of the parameter space is in the fast cooling (upper panels), 13 $\%$ (7 $\%$)transition case (middle panels), while 59 $\%$  (88 $\%$) of the parameter space is in the slow cooling case (lower panels). These values are calculated as the total number of cases in a cooling case over the total number of cases considered.

Furthermore, values of $\varepsilon_{\rm B}$ and $\varepsilon_{\rm e}$ greater than $\gtrsim 10^{-2}$ and $\gtrsim 10^{-1}$, respectively, are strongly preferred for the fast cooling and transition cases while values less than $\lesssim 10^{-2}$ and $\lesssim 10^{-1}$, respectively, correspond to slow cooling cases. In fast cooling and transition cases,  the initial parameter space (without considering the VHE flux UL) is reduced as the keV-fluence decreases and the opposite happens for the slow cooling regime. Dashed lines in Figure~\ref{fig:Histograms} assume a VHE flux UL at 500 GeV of ${\rm 10^{-10} erg\,cm^{-2}}$. The parameter space remaining (dashed lines) after requiring consistency of the light curves with the VHE flux ULs for each time window is strongly reduced for higher keV-fluences while the effect is negligible for the lowest keV-fluence independent of the cooling case. In summary, this analysis will yield a stronger restriction of the parameter space for close GRBs \citep[$z\lesssim 0.3$;][]{2014ARA&A..52...43B} with high keV-fluence. However, since the parameter space in the transition case is similar or smaller compared to the fast cooling case and the results are
similar, hereafter we only present the results for the fast and the slow cooling cases. Furthermore, we exclude results from the two bursts (GRB 170709334 and GRB 180103090) with  keV-fluence below $1\times10^{-6}\rm erg\, cm^{-2}$ since this methodology restricts poorly the parameter space.

\begin{figure}[ht!]
\plotone{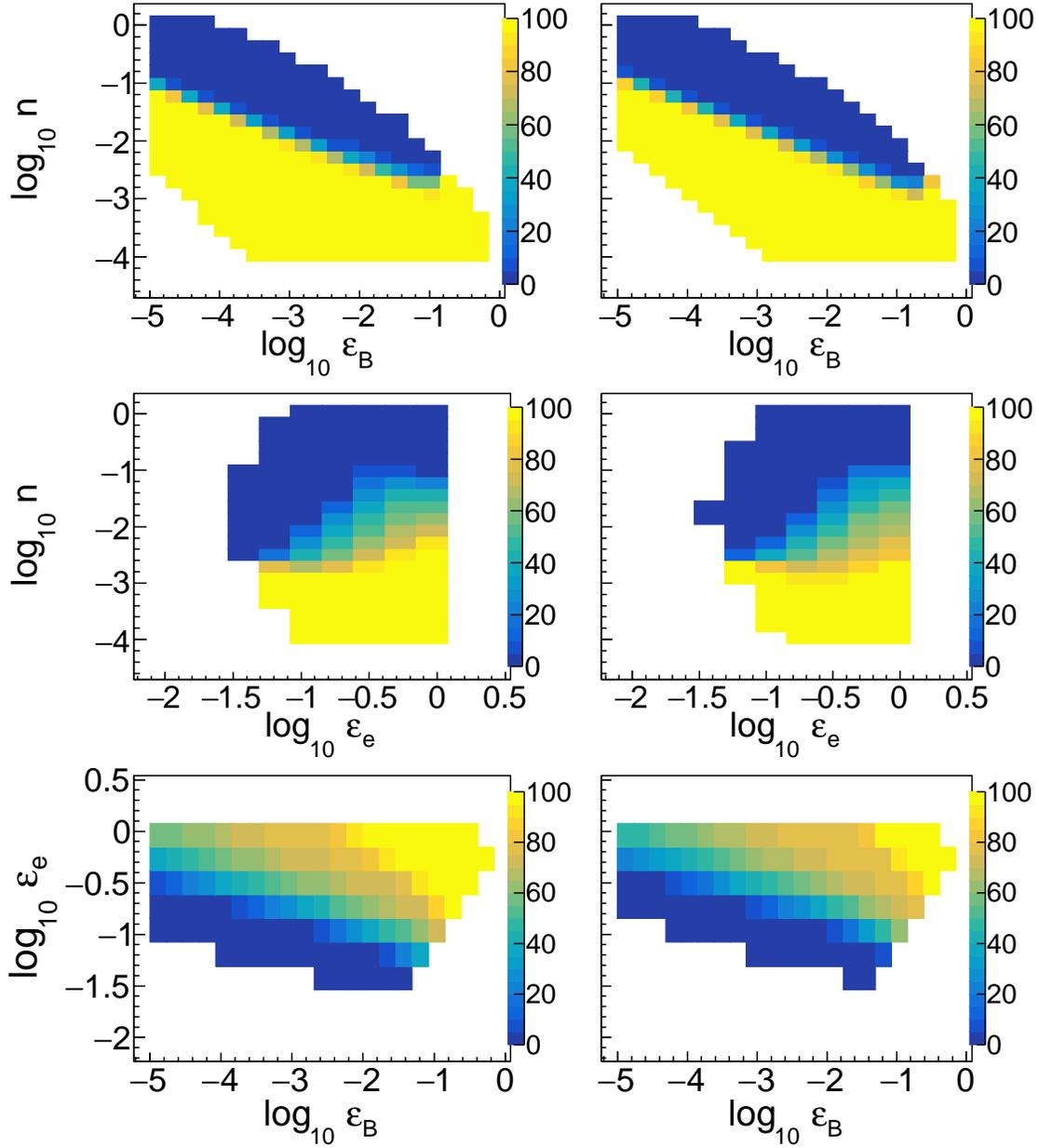}
\caption{Allowed density of the surrounding medium as function of microphysical parameters related with the energy given to amplify the magnetic field (top panels) and to accelerate electrons (middle panels) and, the energy given to accelerate electrons as function of the energy to amplify the magnetic field (botton panels), from left to right columns for GRB 181222A and GRB 170206A. Color scale indicates the percentage of cases remaining from the total cases per bin in the fast cooling regime.
\label{fig:ResFour}}
\end{figure}

\begin{figure}[ht!]
\plotone{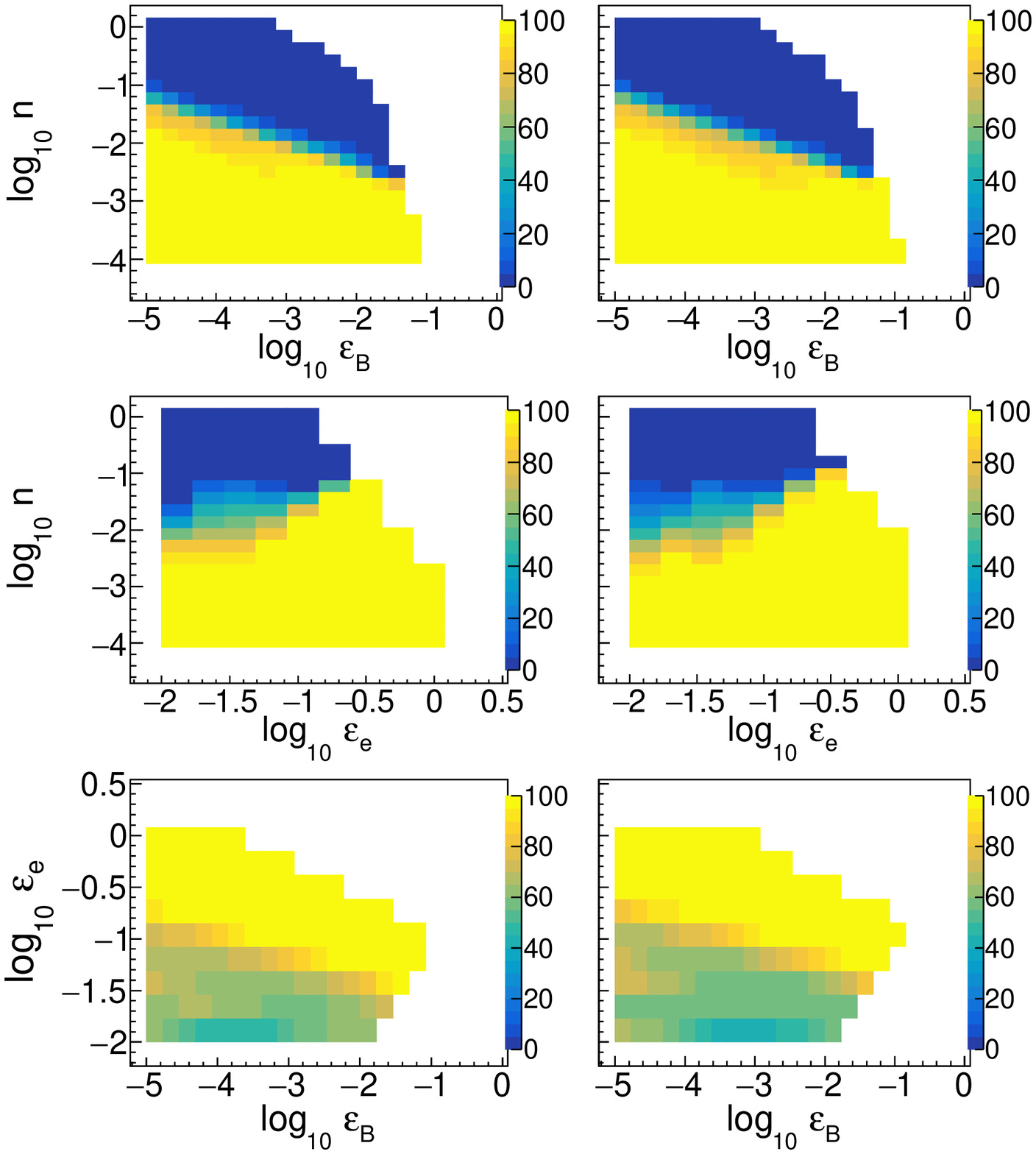}
\caption{Same as Fig. \ref{fig:ResFour}, but for the slow cooling regime.
\label{fig:ResFour-2}}
\end{figure}

Figure~\ref{fig:ResFour} shows results for GRB 181222841 and GRB 170206A for the fast cooling case, considering the corresponding observed keV-fluence, assuming a redshift of $0.3$ and, requiring consistency of the theoretical light curves with HAWC flux ULs at 800 GeV in every time window. As discussed before, the considered parameter space reduces as the keV-fluence decreases. This is observed as a smaller colored area from highest (left panels) to lowest (right panels) keV-fluences.  As the density of the surrounding medium decreases, a higher amount of energy to amplify the magnetic field is required and a dependency is observed (top panels). Thus, for a given value of $\varepsilon_{B}$ a value of $\log_{10}(n)\lesssim 0.5 \log_{10} (\varepsilon_{B}) - 3$ is allowed. As consistency with the HAWC ULs is required, cases with ${\rm n\geq 10^{-2}\,{cm^{-3}}}$ are strongly excluded, see bluish area in Figure~\ref{fig:ResFour} two middle panels. Those cases with  ${\rm n\geq 10^{-2}\,{cm^{-3}}}$ prefer low values of $\varepsilon_B < 10^{-1.5}$. However, as observed from the third line of Figure~\ref{fig:ResFour}, higher values of $\varepsilon_{B}$ and $\varepsilon_{e}$ are strongly preferred in the fast cooling case. Lower limits for the value of the Lorentz factor are calculated using eqn. \ref{Gamma}. In particular for bursts GRB 170206A and GRB 181222841, the bulk Lorentz factors must be larger than 1080 and 1266 respectively. These values are a consequence of requiring that the VHE emission appears within the first 20 s after the trigger time and stays for that long period in a fast cooling regime. For instance, to model the emission observed above 100 MeV up to 100 s in the bright and hard short burst GRB 090510 \citep{2010ApJ...716.1178A}, large values between 1900 and 9000 for the bulk Lorentz factor were found accompanied by values of the medium density as lower as ($\rm n \sim 10^{-6}-10^{-3} cm^{-3}$). These values for $\rm n $ are even lower than the ones reported in this work \citep{2011ApJ...733...22H}.

Figure~\ref{fig:ResFour-2} is the same as Figure~\ref{fig:ResFour} for the slow cooling case. Opposite to the fast cooling case, the considered parameter space reduces as the keV-fluence increases. Again, as the density of the surrounding medium decreases, a higher amount of energy to amplify the magnetic field is required and a dependency is observed, $\log_{10}(n)\lesssim  - 0.4 \log_{10} (\varepsilon_{B}) - 3.4$. For GRB 181222841 and GRB 170206A, as consistency with the HAWC ULs is required, cases with ${\rm n\lesssim 10^{-2}\,{cm^{-3}}}$ are generally preferred, with the upper bound dropping towards $\rm {10^{-3}\,{cm^{-3}}}$ as $\varepsilon_{B}$ increases. Those cases with  ${\rm n\geq 10^{-2}\,{cm^{-3}}}$ prefer values of $\varepsilon_e \gtrsim 10^{-1}$ and therefore $\varepsilon_B \lesssim 10^{-3}$. The corresponding minimum values of the bulk Lorentz factors for bursts GRB 170206A and GRB 181222841, are same as for the fast cooling case when the preferred value of ${\rm n\lesssim 10^{-2}\,{cm^{-3}}}$ are considered. These values are unexpectedly large for slow cooling regime.

The maximum value of $n$ obtained from this analysis is consistent with the evidence that short bursts explode in very low densities, even lower than those reported before by \cite{2006ApJ...650..261S, 2014ARA&A..52...43B}. However, the maximum value of ${\rm n} \lesssim 10^{-3}\,{\rm cm^{-3}}$, could be rejecting the possibility of VHE emission mainly in the fast cooling regime for high keV-fluence bursts since almost no short burst has been observed with such a low density value. However, the joint detection and modeling of two messengers, gravitational and electromagnetic waves \citep{PhysRevLett.119.161101,2017GCN.21520....1V}, of a fusion of two neutron stars \citep{2017ApJ...848L..13A} and its associated short burst provided values of circumburst densities even lower than the limit obtained here. 

\section{Conclusions}
\label{sec:conclusions}
We use data collected by the HAWC gamma-ray observatory to search for VHE emission from a sample of 47 short GRBs detected inside its field of view from December 1st, 2014 and May 14, 2020. Our analysis is oriented to the search for possible delayed or extended VHE emission,  inspecting the signal up to  10 times the duration of the prompt emission. Although no evidence of transient signal is found, we derive fluence upper limits assuming a spectral hypothesis and the theoretical model described by \cite{2001ApJ...548..787S}. We show that by restricting the flux predicted for the SSC emission in the framework of the external shock model, for close and short GRBs with high fluence this analysis could extract information about the ISM density, the bulk Lorentz factor of the expanding blast wave and the microphysical parameters of the fireball.

The most constraining results, assuming a redshift of 0.3, are obtained for bursts with the highest keV-fluences, GRB 170206A and GRB 181222841. For the fast cooling case, we have found that ${\rm n\leq 10^{-2}\,{cm^{-3}}}$, $\varepsilon_{B} \gtrsim 10^{-1.5}$ and, $\varepsilon_{e} \gtrsim 10^{-0.5}$ and $\Gamma\approx 1080$ are required to observe VHE emission for high keV-fluence bursts. These low values of $\rm n$ are consistent with those obtained after modeling  high-energy emission in other bursts (e.g., GRB 090510). Furthermore, there is some evidence of such a low  ${\rm n}$, in particular, interestingly, for the gravitational wave GW1700817 associated to the GRB 170817A. Similar low values for ${\rm n}$ are preferred for the slow cooling case although higher values are not totally excluded if $\varepsilon_{B} \lesssim 10^{-3}$ and, $\varepsilon_{e} \gtrsim 10^{-1}$. We show that it is possible to restrict the microphysical parameters of the SSC forward-shock model even when considering a conservative value of $z=0.3$. 

Two interesting events are coming that may be convenient for an analysis as presented here. First, the new HAWC analysis framework that is more sensitive to photons with energy below a few TeVs. And second, the operation of LIGO and VIRGO, that may give us exciting candidates of close GRBs, possibly with interstellar density values convenient for VHE emission in the fast cooling regime.

\section*{Acknowledgments}
We acknowledge the support from: the US National Science Foundation (NSF); the US Department of Energy Office of High-Energy Physics; the Laboratory Directed Research and Development (LDRD) program of Los Alamos National Laboratory; Consejo Nacional de Ciencia y Tecnolog\'ia (CONACyT), M\'exico, grants 271051, 232656, 260378, 179588, 254964, 258865, 243290, 132197, A1-S-46288, A1-S-22784, c\'atedras 873, 1563, 341, 323, Red HAWC, M\'exico; DGAPA-UNAM grants IG101320, IN111716-3, IN111419, IA102019, IN110621, IN110521; VIEP-BUAP; PIFI 2012, 2013, PROFOCIE 2014, 2015; the University of Wisconsin Alumni Research Foundation; the Institute of Geophysics, Planetary Physics, and Signatures at Los Alamos National Laboratory; Polish Science Centre grant, DEC-2017/27/B/ST9/02272; Coordinaci\'on de la Investigaci\'on Cient\'ifica de la Universidad Michoacana; Royal Society - Newton Advanced Fellowship 180385; Generalitat Valenciana, grant CIDEGENT/2018/034; The Program Management Unit for Human Resources; Institutional Development, Research and Innovation, NXPO (grant number B16F630069); Coordinaci\'on General Acad\'emica e Innovaci\'on (CGAI-UdeG), PRODEP-SEP UDG-CA-499; Institute of Cosmic Ray Research (ICRR), University of Tokyo, H.F. acknowledges support by NASA under award number 80GSFC21M0002. We also acknowledge the significant contributions over many years of Stefan Westerhoff, Gaurang Yodh and Arnulfo Zepeda Dominguez, all deceased members of the HAWC collaboration. Thanks to Scott Delay, Luciano D\'iaz and Eduardo Murrieta for technical support.

\bibliography{references}{}
\bibliographystyle{aasjournal}

\begin{deluxetable*}{l|cccccccccc}
\renewcommand{\arraystretch}{1.0}\addtolength{\tabcolsep}{-2pt}
\tabletypesize{\footnotesize}
\tablecolumns{11}
\tablewidth{0pt}
\tablecaption{Short GRBs detected in the field of view of HAWC}
\tablehead{\colhead{GRB name} & \colhead{Detection Time}   & \colhead{R.A. J2000} & \colhead{Dec. J2000} & \colhead{Error Box} & \colhead{Zenith angle} & \colhead{$T_{90}$} &\colhead{R.A} & \colhead{Dec.}& \colhead{HAWC UL } & \colhead{Significance}\\
	   \vspace{-0.1cm} & \colhead{(UT)} & \colhead{(deg)} & \colhead{(deg)} & \colhead{} 
                           & \colhead{(deg)} & \colhead{(s)}& \colhead{(deg)} & \colhead{(deg)} &\colhead{[0-2s](${\rm erg\,cm^{-2}\,s^{-1}}$)} & \colhead{($\sigma$)}}
\startdata
 & \multicolumn{6}{c}{\it Fermi-GBM}  \\ 
\hline
GRB 141202470  &  11:17:05.606 & 145.01 & 59.87  & $3.32^{\circ}$ & 40.83 & 1.34  & 142.89 & 60.7 & 1.603e-05 & 0.43 \\
GRB 150201040  &  00:56:54.289 & 5.63 &	19.75 & $13.92^{\circ}$ & 39.53 & 0.51  &11.28 & 8.13 & 8.420e-06 & -1.82 \\
GRB 150522944  &  22:38:44.068 &130.86&	58.58 & $10.47^{\circ}$ & 40.02 & 1.02  & 112.49 & 55.82 &8.517e-06 & -0.52 \\
GRB 150705588  &  14:07:11.608 & 66.54 & -6.62 & $12.60^{\circ}$ & 38.34 & 0.70 &63.12 & 3.31 &2.767e-06 & 0.57 \\
GRB 150811849  &  20:22:13.749 &186.35&-14.11 & $0.99^{\circ}$ & 37.77 & 0.64  &- &- &2.512e-06  & -- \\
GRB 150906944  &  22:38:47.307 & 212.04 & 1.09  & $5.19^{\circ}$ & 23.76 & 0.32   &209.83 &5.52 &1.333e-06 & 0.34\\
GRB 150923297  &  07:07:36.184 &316.80&31.82  & $10.76^{\circ}$ & 50.29 & 0.19  & 327.262 & 38.1856 & 2.926e-05 & -0.49\\
GRB 151022577  &  13:51:02.089 &110.37&	40.23  & $21.36^{\circ}$ & 33.64 & 0.32 &120.27 & 49.19 & 6.415e-06  &-0.86\\
GRB 160211119  &  02:50:48.276 & 123.20 & 53.43 & $4.97^{\circ}$ & 44.87 & 0.96  &121.57 & 51.57 & 2.921e-05 &0.02\\
\textbf{GRB 160406503}  &  12:04:36.798 & 261.80 & 32.26 & $11.80^{\circ}$ & 20.28 & 0.43 & 266.8 & 22.05 & 7.265e-07 &-0.03\\
GRB 160820496  &  11:54:10.646 &6.82 & 21.85  & $4.52^{\circ}$ & 40.67 & 0.38  &- &- & 2.542e-06 & -\\
GRB 161026373  &  08:57:16.281 &67.70 & 41.85& $11.68^{\circ}$ & 23.16 & 0.11  &54.67 & 37.98 & 1.968e-06  & 0.65\\
GRB 170203486  &  11:40:25.855 & 245.09 &-0.51 & $14.06^{\circ}$ & 38.39 & 0.34 &234.97 &-9.55 & 1.157e-05  & 1.85\\
GRB 170219002  &  00:03:07.123 & 54.84 & 50.07 & $1.41^{\circ}$ & 31.20 & 0.10  &54.19 &51.32 & 2.179e-06 &-1.03 \\
GRB 170403583  &  13:59:17.798 &267.08 & 14.53& $7.16^{\circ}$ & 35.96 & 0.48  &- &- & 1.133e-05 & -\\
GRB 170604603  &  14:28:05.086 &340.44 & 40.67 & $4.10^{\circ}$ & 35.12 & 0.32  &344.3 & 42.56 & 4.005e-06  & 0.03 \\
\textbf{GRB 170709334}  &  08:00:23.979 & 310.04 & 2.20  & $7.49^{\circ}$ & 16.73 & 1.86 &311.76 & 5.563 & 6.360e-07  & -1.65\\
GRB 170818137  &  03:17:19.979 & 297.22 & 6.35  & $11.54^{\circ}$ & 22.16 & 0.58 &287.11 & 9.09 & 8.255e-07 & 0.18 \\
GRB 170826369  &  08:51:07.514 & 64.34 & 21.07& $0.86^{\circ}$ & 50.83 & 0.26 &- &- & 2.355e-04  &--\\
GRB 171011810  &  19:26:27.946 & 177.04 & 26.93& $15.96^{\circ}$ & 35.39 & 0.48 & 187.80 &14.53 & 2.150e-06  &-0.75 \\
GRB 171207055  &  01:18:42.452 & 314.39 & 51.67& $9.54^{\circ}$ & 47.33 & 0.18  & 319.53 & 48.74 & 2.935e-05 & -0.49\\
\textbf{GRB 180103090}  &  02:09:12.118 & 25.40 & 28.01  & $6.97^{\circ}$ & 14.26 & 0.02 & 32.51 & 28.99 & 7.235e-07 & 0.59\\
GRB 180128881  &  21:09:19.457 & 323.18 & -13.55 & $9.22^{\circ}$ & 40.56 & 1.79  &314.86 &-11.37 & 6.350e-05 & 1.14\\
\textbf{GRB 181222841}  & 20:11:37.438 & 311.15 & 22.86 & $1.60^{\circ}$ & $14.03$ & $0.576$ & $311.68$ & $22.63$ & 4.935e-07 &$-0.782$\\
GRB 190226515 &   12:21:45.676 & 224.43 & -8.61 &    $5.11^{\circ}$   &      33.71 &	  0.192	 & 221.914 &-12.88 & 3.397e-06 &-1.083\\
GRB 190630257 &   06:09:58.319 & 306.98 & -1.33 &    $7.65^{\circ}$   &      39.07 &	  0.224	 & 312.63 & -0.36 & 1.231e-05 & -0.824\\
GRB 190724031 &   00:43:56.792 & 170.35 & 15.15&    $13.62^{\circ}$   &      42.66 &	  0.08 & 166.99 & 5.44 & 6.350e-05 & -0.177 \\
\textbf{GRB 190905985} &   23:38:28.489 & 234.48 & 3.12 &    $3.62^{\circ}$   &      17.49 &	  0.704	& 232.32 & 0.36 & 7.925e-07 & -1.577\\
GRB 191031891 &   21:23:31.128 & 283.27 & 47.64 &    $0.05^{\circ}$   &      32.94 &	  0.256	 & - & - & 1.138e-06 & -\\
GRB 191117637 &   15:17:38.361 & 157.42 & 7.23 &    $13.07^{\circ}$   &      32.21 &	  1.28	 & 152.56 & -3.97 & 3.697e-05 &-0.105\\
GRB 200221162&   03:52:58.711  & 157.10 & 33.14 &    $5.04^{\circ}$   &    43.20 &	  1.728	 &153.69 & 35.74 & 1.605e-05 &-0.4236\\
GRB 200224416 &  09:58:44.567  & 187.02 & -19.55 & $12.52^{\circ}$ & 42.85 & 0.064  &187.28 &-13.06 & 8.470e-06 & -0.865\\
GRB 200423579 & 13:54:11.373   & 325.02 & 66.78  & $11.20^{\circ}$ & 47.88 & 0.032 & 312.924 & 57.27 & 9.970e-06 & -1.635  \\
\textbf{GRB 200514380} & 09:07:37.124 & 238.32 & 37.22 & $13.12^{\circ}$ & 34.52 & 1.664  & 242.86 & 34.16 & 3.245e-06 & -1.077\\
\hline
 & \multicolumn{6}{c}{{\it Swift}-BAT} \\ 
\hline
\\
\textbf{GRB 141205A}  &  08:05:17 & 92.86  & 37.88 & $0.03^{\circ}$ & 19.36 & 1.1 &- &- & 4.829e-07 & -  \\
\textbf{GRB 150423A}  &  06:28:04 & 221.58 & 12.28 &  $0.0004^{\circ}$ & 12.63 & 0.22 &- &- &2.163e-07 & - \\
\textbf{GRB 150710A}  &  00:28:02 & 194.47 & 14.32 & $0.0008^{\circ}$ & 5.40 & 0.15 &- &- & 4.630e-07 & - \\
GRB 160714A  &  02:19:15 & 234.49 & 63.81 & $0.045^{\circ}$ & 44.88 & 0.35 &- &- &2.041e-05 & -- \\
GRB 170112A  &  02:02:00 & 15.23 & -17.23 & $0.042^{\circ}$ & 46.34 & 0.06 &- &- & 4.66e-05 & -- \\
\hline
 & \multicolumn{6}{c}{\it Fermi - LAT} \\ 
\hline
\\
\textbf{GRB 170206A} & 10:51:57.696 & 212.79 & 14.48 & $0.85^{\circ}$ & 11.10 & 1.168 &- &- & 4.736e-07 & - \\
GRB 180225417 &   10:00:54.175  & 180.98 & -9.49  &   $7.47^{\circ}$     &   39.01   &  0.896 &178.44 &-13.19 & 3.710e-05 & 1.111  \\
GRB 180402406 &   09:44:59.367  &  251.90 & -14.96 &    $0.05^{\circ}$    &    36.16     &   0.448 & - & - & 2.470e-06 &-- \\
GRB 180511364 &   08:43:35.786  &	250.42 & -8.18 &    $15.07^{\circ}$  &      29.67 &	  0.128 & 26.68 & 247.96 & 2.156 e-06 & -0.769\\
GRB 180511437 &   10:29:52.606 &	257.78 & 9.07 &    $10.16^{\circ}$    &    31.91	&  1.984 &260.16 & 14.15 & 2.745 e-06 & 0.378    \\  
\textbf{GRB 180617872} &   20:55:23.463 &106.89 & 24.87 &    $8.24^{\circ}$      &   15.45	&  1.920 & 106.50 & 25.74 & 1.052e-06 & 0.868\\
GRB 180626392  &  09:23:50.648 & 285.06 & 44.82 &    $8.21^{\circ}$   &      37.59	&  0.960 & 294.37 & 46.65 & 4.691e-06 & 0.575 \\
GRB 180803590  &  14:09:49.734 & 71.63 & 57.65 &    $17.37^{\circ}$ &      38.84 &	  0.384 &68.736 &56.7804 & 8.505e-06 & -1.83\\
 \enddata
\tablecomments{Observational information for bursts detected in the field of view of HAWC from December 2014 to May 2020 by {\it Fermi}-GBM, Swift/BAT and {\it Fermi}-LAT. {\bf The 3th and 4th columns show the GRB position as reported by  {\it Fermi}-GBM, Swift/BAT or {\it Fermi}-LAT}.  The 6th column gives the angle between the GRB position and the HAWC zenith. {\bf The 8th and 9th columns state the position of the one-degree circle where the maximum significance is observed for the first 2 s after the trigger}. We do not report the uncertainty for GRBs with an X-ray or optical counterpart. The HAWC upper limits to the flux are given assuming an spectral index of -0.5 as in \cite{2017ApJ...843...88A}. }
\label{tbl:positions}
\end{deluxetable*}

\begin{deluxetable*}{ccccc}
\tablenum{2}
\tablecolumns{4}
\tablecaption{Information of time windows with significance greater than 2$\sigma$}
\label{tbl:HighSignificance}
\tablewidth{0pt}
\tablehead{
\colhead{GRB name } & \colhead{One-degree circle position} &\colhead{Time Bin} & \colhead{Significance} & HAWC UL\\
& \colhead{(RA,Dec)deg} & (s) &\colhead{$\sigma$} & \colhead{(${\rm erg\,cm^{-2}\,s^{-1}}$)} 
}
\startdata
GRB 161026373  & (74.24 , 43.98)  & 2-4  & 2.53 & 1.926e-06 \\
GRB 171207055  & (327.95,52.94)  & 18-20 &  2.09 & 2.375e-05 \\
GRB 180803590  & (67.91 , 48.98)  & 4-6  & 2.31& 3.527e-06  \\
GRB 181222841  & (311.68, 21.43)  & 14-16 & 2.13 & 8.259e-07 \\
GRB 200423579  & (330.62 , 60.57) & 8-10 & 2.03& 2.379e-05    \\
GRB 200514380  & (225.24, 33.26)  & 14-16  & 3.16 & 8.583e-05 \\
\enddata
\tablecomments{The HAWC upper limits to the flux are given assuming an spectral index of -0.5 as in \cite{2017ApJ...843...88A} }
\end{deluxetable*}


\begin{deluxetable*}{cchlDlc}
\tablenum{3}
\tablecolumns{6}
\tablecaption{Relevant information of analyzed bursts}
\label{tbl:fourgrbs}
\tablewidth{0pt}
\tablehead{
\colhead{GRB name } & \colhead{Detection Time} & \nocolhead{Common} & \colhead{$E_{iso}$} &
\multicolumn2c{Zenith angle} & \colhead{References} & \colhead{} \\
\colhead{} & \colhead{(UT)} & \nocolhead{Name} & \colhead{(erg)} &
\multicolumn2c{(deg)} & \colhead{} & \colhead{}
}
\startdata
GRB 181222A & 20:11:37.438 & Messier 1 & $8.199\times10^{51}$ & 14.0 &  &\citet{2018GCN.23548....1V} \\
GRB 170206A & 10:51:57.696 & Messier 2 & $2.297\times10^{51}$ & 11.1 &  & \citet{2017GCN.20616....1V} \\
GRB 170709A & 08:00:23.979 & Messier 3 & $1.55\times10^{50}$ & 16.73 &  &   \cite{2020ApJ...893...46V} \\
GRB 180103A & 02:09:12.118 & Messier 4 & $2.010\times10^{49}$ & 14.3 & & \citet{2018GCN.22305....1B} \\
\enddata
\end{deluxetable*}

\clearpage
\appendix
\section{A. Coefficients of the synchrotron-self Compton light curves}
The coefficients for the break energies are
\bary
A_{\rm \gamma,m}&=&\frac{3^{3/4}\,q_e\,m_p^{15/4} (p-2)^4}{2^{1/4}\,\pi^{1/4}\, m_e^5 (p-1)^4}\cr
A_{\rm \gamma,c}&=&\frac{3^{11/4}\,\pi^{7/4}\,q_e\,m_e^3}{2^{25/4}\,m_p^{9/4}\,\sigma_T^4}\cr
A^{\rm KN}_{\rm \gamma,c}&=&\frac{3^\frac23\,\pi^\frac13\,m_e^2}{2^\frac73\,\sigma_T\, m_p^\frac23}\cr
F_{\rm \gamma,max}&=&\frac{m_e\, \sigma_T^2}{2^{7/4}\,3^{3/4}\,\pi^{3/4}\,m^{3/4}_p\,q_e }
\eary
and for the light curves in fast- and slow- cooling regimes are
{\small
\bary\nonumber
A_{\rm f1}&\simeq& \frac{2^{\frac13}\,\sigma_T^{\frac{10}{3}}}{3^{\frac53}\,7^{\frac{3}{12}}\,\pi^{\frac34}\,q_e^{\frac43}}\,(1+Y)^{\frac43}\,(1+z)\, \varepsilon_{B}^\frac53 \,n^2\,D^{-2}\,E^\frac53\,,\cr
A_{\rm f2}&\simeq& \frac{3^{\frac58} \pi^{\frac18} m_e^{\frac52}}{2^{\frac{39}{8}}\,q_e^{\frac12}\,m_p^{\frac{15}{8}}} \,(1+Y)^{-2}  \,(1+z)^\frac38\, \varepsilon_{B}^{-\frac{5}{4}}\,D^{-2}\,n^\frac18\,E^\frac58\,,\cr
A_{\rm f3} &\simeq& \frac{3^{\frac{2-3p}{8}} m^{\frac{10-5p}{2}} _e\,q_e^{p-2} (p-1)^{2-2p}} {2^{\frac{38+p}{8}} m^{\frac{30-15p}{8}}_p \pi^{\frac{6+5p}{8}}  (p-2)^{2-2p}}  (1+Y)^{-2} (1+z)^\frac{5p-2}{8}\,\varepsilon_{B}^\frac{p-6}{8}\,\varepsilon_{e}^{2p-2}\,D^{-2}\,E^\frac{3p+2}{8}\,, \cr
A_{\rm s1}&\simeq&\frac{m_e^\frac83\,\sigma_T^2 (p-1)^\frac43}{2^\frac53 3\,\pi^\frac23 \,q_e^{\frac43}\,m_p^2\,(p-2)^{\frac43}} \,(1+z)^\frac13\,\varepsilon_{B}^{\frac13}\,\varepsilon_{e}^{-\frac{4}{3}}\,n^\frac43\, D^{-2}\,E\,,\cr
A_{\rm s2}&\simeq&\frac{3^{\frac{3p-9}{8}}\,m_p^\frac{15p-21}{8}\sigma_T^2\,q_e^\frac{p-3}{2}   (p-2)^{2(p-1)}} {2^\frac{p-6}{8}\pi^\frac{p+5}{8} m_e^\frac{5p-9}{4} (p-1)^{2(p-1)}}(1+z)^\frac{5p+1}{8}\,\varepsilon_{B}^\frac{p+1}{4}\,\varepsilon_{e}^{2(p-1)}\,n^{\frac{11-p}{8}}\,D^{-2}\,E^\frac{3p+7}{8}\,,\cr
A_{\rm s3}&\simeq&\frac{3^{\frac{3p+2}{8}}\,m_p^\frac{15p-30}{8}\,q_e^\frac{p-2}{2} (p-1)^{2(p-1)}} {2^\frac{p+38}{8}\pi^\frac{p-2}{8} m_e^\frac{5p-10}{2} (p-2)^{2(p-1)}}(1+z)^\frac{5p-2}{8}\,t_0^\frac{p-2}{p-4}\, \varepsilon_{B}^\frac{p-2}{4}\,\varepsilon_{e}^{2p-3}\,n^{\frac{2-p}{8}}\,D^{-2}\,E^\frac{3p+2}{8}\,,\cr
\eary
}
where $m_p$ in the proton mass, $m_e$ is the electron mass, $q_e$ is the elementary charge and $\sigma_T$ is Thompson cross section. The sub-indexes ${\rm f}$ and ${\rm s}$ refer to fast and slow-cooling regime, respectively, and ${\rm h}$, ${\rm m}$ and ${\rm l}$ to high, medium and low power laws.
\label{appendix}

\end{document}